\documentclass[10pt]{iopart}

\usepackage{graphicx}
\usepackage{iopams}

\begin{document}

\title{Hubbard-Kanamori model: spectral functions, negative electron compressibility, and susceptibilities}

\author{A Sherman}

\address{Institute of Physics, University of Tartu, W. Ostwaldi Str 1, 50411 Tartu, Estonia}

\ead{alekseis@ut.ee}

\begin{abstract}
The two-orbital Hubbard-Kanamori model is studied using the strong coupling diagram technique. This approach allows one to take into account the interactions of electrons with spin, charge, and orbital fluctuations of all ranges. It was found that, at low temperatures, the model has four regions of the negative electron compressibility, which can lead to charge inhomogeneities with the assistance of phonons. For half-filling, the phase diagram of the model contains regions of the Mott and Slater insulators, bad-metal, and states with spin-polaron peaks. These sharp peaks at the Fermi level are seen in doped states also. A finite Hund coupling leads to a large increase of antiferromagnetic spin correlations with strong suppression of charge and orbital fluctuations near half-filling. For moderate doping, all types of correlations become comparable. They peak at the antiferromagnetic wave vector except for a narrow region with an incommensurate response. Stronger doping destroys all correlations.
\end{abstract}

\vspace{2pc}
\noindent{\it Keywords}: Hubbard-Kanamori model, strong coupling diagram technique, charge instabilities, phase diagram, spin polarons, susceptibilities


\maketitle

\ioptwocol

\section{Introduction}
The Hubbard-Kanamori model (HK) \cite{Kanamori,Georges13} is a ge\-ne\-ra\-li\-za\-tion of the Hubbard model to a many-band case, which takes into account the Hund coupling. The model is used for the description of crystals with partially filled $d$- and $f$-shells, which are characterized by moderate to strong on-site Coulomb repulsions. Among these crystals are transition metal oxides, ruthenates, iron pnictides and chalcogenides, which exhibit several bands crossing the Fermi level with a pronounced Hund's coupling. Another peculiarity of these crystals is the phase separation (see, e.g., \cite{Shenoy,Dai,Lang}), which is ubiquitous for these classes of crystals, as well as for another group of strongly correlated materials -- cuprate perovskites \cite{Sigmund}. A frequently used approach for investigating the HK model is the dynamical mean-field theory (DMFT) \cite{Georges13,Georges96,Stadler}. However, as indicated in \cite{Stadler}, this approximation does not describe the phase separation. Such a state was obtained using another approach -- the slave-spin mean-field approximation -- in \cite{Medici}.

In the case of strong correlations, it is reasonable to apply the perturbation series expansion around the atomic limit. In this work, we use such an expansion termed the strong coupling diagram technique (SCDT) \cite{Vladimir,Metzner,Pairault,Sherman18,Sherman19a} for investigating the HK model. The approach is based on a regular series expansion of Green's functions in powers of hopping constants of the kinetic term of the Ha\-mil\-to\-ni\-an. Terms of the series are products of hop\-ping constants and on-site cumulants of electron operators. The linked-cluster theorem is valid, and par\-tial summations are allowed in this diagram technique (the concise description of the approach can be found in \cite{Sherman16}). For the two-dimensional (2D) one-band Hubbard model, the validity of the SCDT was proved in numerous comparisons with the results of numerical experiments and experiments with ultracold fermionic atoms in 2D optical lattices. In particular, it was shown that the critical repulsion for the Mott metal-insulator transition is close to that observed in Monte-Carlo simulations. For the comparable parameters, spectral functions and den\-si\-ties of states are similar to those found in ex\-act diagonalizations and Monte Carlo simulations \cite{Sherman18}. Temperature and concentration dependencies of the uniform spin susceptibility, spin structure fac\-tor, square of the site spin, and double occupancy are in good agreement with re\-sults of Monte Carlo simulations, numeric linked-cluster expansion, and experiments with ul\-tra\-cold fermionic atoms \cite{Sherman18,Sherman19a}. Shapes and intensity distributions in Fermi surfaces in electron- and hole-doped cases are similar to those observed experimentally \cite{Sherman18,Sherman19b}. Lastly, moments sum rules are fulfilled with good accuracy \cite{Sherman18}. These facts give grounds to believe that this approach will be equally useful for investigating many-band Hubbard models. Such a generalization was carried out in \cite{Sherman20a} for the Emery model describing electrons in Cu-O planes of cuprates. In particular, it was shown that the magnetic response is commensurate in the electron-doped case and incommensurate in the hole-doped case, the correlations remain strong in the former case and weaken rapidly in the latter. These results reproduce correctly experimental observations \cite{Damascelli,Armitage}.

The SCDT allows one to take into account interactions with spin, charge, and, in the present model, orbital fluctuations. In the ladder approximation used in this work, these interactions are described by diagrams with ladder inserts. If these ladders are constructed from renormalized hopping lines and second-order cumulants as irreducible four-leg vertices, ladders of all lengths can be summed. Thereby, fluctuations of all ranges are taken into account in an infinite crystal.

In this work, we consider the HK model containing two degenerate orbitals on a 2D square lattice. The small number of orbitals was chosen to reduce the computational effort and sizes of arrays. The range of on-site Coulomb repulsions $U$ from $2t$ to $8t$ and two values of Hund's coupling $J=U/4$ and $J=0$ are investigated. Here $t$ is the hopping constant between nearest-neighbor sites. Derived equations for the electron Green's function are self-consistently solved for the range of temperatures $T$ from $0.07t$ to $0.9t$. For low temperatures, the calculated dependencies of the electron concentration $x$ on the chemical po\-ten\-ti\-al $\mu$ demonstrate four regions of the negative electron compressibility (NEC) $\kappa=x^{-2}({\rm d}x/{\rm d}\mu)<0$. These regions arise due to level crossing in the site Hamiltonian occurring at certain values of $\mu$. As a consequence, drastic changes in electron energy bands proceed near these chemical potentials. In these regions, the electron concentration decreases as the chemical potential grows. Such behavior of the compressibility gives rise to charge instability with the assistance of phonons. Sharp band changes are reflected in densities of states (DOS) as well as in magnetic, charge, and orbital susceptibilities. According to the Hund rule, the combined spins $S=1$ appear near half-filling for $J=U/4$. This leads to a large growth of antiferromagnetic spin correlations and significant suppression of orbital fluctuations in these conditions. For the cases $J=0$ and $J=U/4$, we investigate the phase diagram at half-filling also. In the considered ranges of parameters, in both cases, phase diagrams contain regions of Mott and Slater insulators, bad metal, and states with the pronounced spin-polaron peak. This peak is a manifestation of bound states of electrons with spin excitations \cite{Sherman19a,Sherman19b}. It is observed both at half-filling for moderate Coulomb repulsions and temperatures, and, in wider ranges of parameters, in the case of doping.

The paper is organized as follows. In section~2, the HK model and the SCDT are introduced. Calculated DOSs are considered in section~3. The origin of NEC regions and their possible consequences are discussed in section~4. Phase diagrams at half-filling are gi\-ven in section~5. Results on the spin, charge, and orbital susceptibilities are presented in section~6. The conclusions of this study are reviewed in section~7.

\section{Model and SCDT method}
The HK Hamiltonian reads \cite{Kanamori,Georges13}
\begin{eqnarray}
H&=&\sum_{{\bf ll'}i\sigma}t_{\bf ll'}a^\dagger_{{\bf l'}i\sigma}a_{{\bf l}i\sigma}+\sum_{\bf l}H_{\bf l},\label{HK}\\
H_{\bf l}&=&\sum_{i\sigma}\bigg[-\mu n_{{\bf l}i\sigma}+\frac{U}{2}n_{{\bf l}i\sigma}n_{{\bf l}i,-\sigma}\nonumber\\
&&+\frac{U-2J}{2}n_{{\bf l}i\sigma}n_{{\bf l},-i,-\sigma}+\frac{U-3J}{2}n_{{\bf l}i\sigma}n_{{\bf l},-i,\sigma}\nonumber\\
&&+\frac{J}{2}\big(a^\dagger_{{\bf l}i\sigma}a^\dagger_{{\bf l}i,-\sigma}a_{{\bf l},-i,-\sigma}a_{{\bf l},-i,\sigma} \nonumber\\
&&\quad\quad-a^\dagger_{{\bf l}i\sigma}a_{{\bf l}i,-\sigma}a^\dagger_{{\bf l},-i,-\sigma}a_{{\bf l},-i,\sigma}\big)\bigg], \label{local}
\end{eqnarray}
where $a^\dagger_{{\bf l}i\sigma}$ and $a_{{\bf l}i\sigma}$ are creation and annihilation operators of electrons on the site {\bf l} of the 2D square lattice in the orbital $i=\pm 1$ with the spin projection $\sigma=\pm 1$, $n_{{\bf l}i\sigma}=a^\dagger_{{\bf l}i\sigma}a_{{\bf l}i\sigma}$ is the occupation-number operator, and $t_{\bf ll'}$ is the hopping integral. In this work, it is orbital-diagonal and nonzero only between nearest-neighbor sites, for which $t_{\bf ll'}=-t$. The Hamiltonian (\ref{HK}), (\ref{local}) is similar to those used in works \cite{Kanamori,Georges13,Stadler,Aron}). In the following, we
consider two values of Hund's coupling -- $J=0$ and $J=U/4$. These values are typical for works on the HK model. For $J=U/4$, all three density-density terms in (\ref{local}) describe repulsive interactions between electrons.

Hamiltonian~(\ref{HK}), (\ref{local}) has the particle-hole symmetry. As a consequence of this, DOSs and spectral functions for a chemical potential $\mu$ are specular reflections of spectra for $\mu'=3U-5J-\mu$ in the vertical line passing through the point $\omega=0$. Due to the same reason the dependence of the electron concentration $x=\sum_{i\sigma}\langle n_{{\bf l}i\sigma}\rangle$ on $\mu$ has the center of symmetry in the point $x=2$, $\mu=\mu_{\rm hf}=(3U-5J)/2$, the chemical potential of half-filling. Therefore, one can limit oneself by the range of chemical potentials $\mu\leq\mu_{\rm hf}$.

We shall calculate the one-particle Green's fun\-cti\-on $G_{i'i}({\bf l'}\tau';{\bf l}\tau)=\langle{\cal T}\bar{a}_{{\bf l'}i'\sigma}(\tau')a_{{\bf l}i\sigma}(\tau)\rangle$, where ${\cal T}$ is the chronological operator, the thermodynamic ave\-ra\-ging and time dependencies are determined by the Hamiltonian (\ref{HK}),
\begin{equation*}
\bar{a}_{{\bf l}i\sigma}(\tau)={\rm e}^{H\tau}a^\dagger_{{\bf l}i\sigma}{\rm e}^{-H\tau}.
\end{equation*}
As mentioned in the introduction, for this purpose, we use the SCDT. Terms of this power expansion are products of hopping integrals $t_{\bf ll'}$ and on-site cumulants \cite{Kubo} of the operators $a^\dagger_{{\bf l}i\sigma}$ and $a_{{\bf l}i\sigma}$. Each term of the expansion can be represented graphically as a diagram, in which hopping integrals are depicted by directed lines and cumulants by circles with the number of outgoing (ingoing) lines corresponding to the cumulant order.

\begin{figure}[t]
\centerline{\resizebox{0.99\columnwidth}{!}{\includegraphics{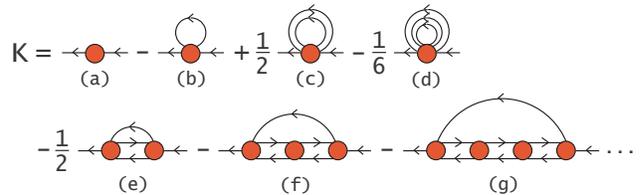}}}
\caption{Diagrams of several lowest orders in the SCDT expansion for ${\bf K}({\bf k},j)$.} \label{Fig1}
\end{figure}

As in the more conventional diagram technique with the power series expansion in interaction \cite{Abrikosov}, the SCDT diagrams can be divided into reducible and irreducible. The latter cannot be separated into two disconnected parts by cutting a hopping line. If the sum of all irreducible diagrams -- the irreducible part -- is denoted by {\bf K}, the Fourier transform of the Green's function can be written as
\begin{equation}\label{Larkin}
{\bf G}({\bf k},j)={\bf K}({\bf k},j)[{\bf 1}-{\bf t_k}{\bf K}({\bf k},j)]^{-1},
\end{equation}
where $j$ is an integer defining the Matsubara frequency $\omega_j=(2j-1)\pi T$, {\bf k} is the 2D wave vector with the components $k_x$ and $k_y$. In Eq.~(\ref{Larkin}), we use matrix notations for quantities depending on two indices $i$; {\bf 1} is the $2\times 2$ unit matrix, and in the present case ${\bf t_k}=-2t\left[(\cos(k_x)+\cos(k_y)\right]{\bf 1}$ (the intersite distance is set as the unit of length). Several lowest order diagrams for ${\bf K}$ are shown in figure~\ref{Fig1}.

Diagrams for ${\bf K}({\bf k},j)$ contain the subset with ladder inserts (see the second row of diagrams in Fig.~\ref{Fig1}). In the ladder approximation, ladder diagrams describe spin, charge, and orbital susceptibilities (see below). Two-leg diagrams for ${\bf G}({\bf k},j)$ with such inserts give an account of the interactions of electrons with the respective excitations. Having regard to such processes of all orders, the irreducible part reads \cite{Sherman20a}
\begin{eqnarray}\label{K}
&&K_{i'i}({\bf k},j)=C^{(1)}_{i'i}(j)+\frac{T^2}{2N}\sum_{{\bf k'}j'\nu}\sum_{i'_1i_1i'_2} \sum_{i_2i'_3i_3}\theta_{i_3i'_3}({\bf k'},j')\nonumber\\
&&\quad\times{\cal T}_{i_1i'_1i_2i'_2}({\bf k-k'},j+\nu,j'+\nu)\nonumber\\
&&\quad\times\bigg[\frac{3}{2}C^{(2)a}_{i'i_1i'_2i_3}(j,j+\nu,j'+\nu,j')\nonumber\\
&&\quad\times C^{(2)a}_{i'_1ii'_3i_2}(j+\nu,j,j',j'+\nu)\nonumber\\
&&\quad+\frac{1}{2}C^{(2)s}_{i'i_1i'_2i_3}(j,j+\nu,j'+\nu,j')\nonumber\\
&&\quad\times C^{(2)s}_{i'_1ii'_3i_2}(j+\nu,j,j',j'+\nu)\bigg] \nonumber\\
&&\quad-\frac{T}{N}\sum_{{\bf k'}j'}\sum_{i_1i'_1}\theta_{i_1i'_1}({\bf k'},j')
\bigg[\frac{3}{2}V^a_{i'ii'_1i_1}({\bf k-k'},j,j,j',j')\nonumber\\
&&\quad+\frac{1}{2}V^s_{i'ii'_1i_1}({\bf k-k'},j,j,j',j')\bigg].
\end{eqnarray}
In this equation, $C^{(1)}_{i'i}(j)$ is the first-order cumulant. Using the possibility of the partial summation in the SCDT, we inserted all possible two-leg diagrams in the internal hopping lines of diagrams in Fig.~\ref{Fig1}. As a result, the bare hopping lines ${\bf t_k}$ are substituted by the renormalized ones described by the expression
\begin{equation}\label{theta}
\mbox{\boldmath $\theta$}({\bf k},j)={\bf t_k}+{\bf t_k}{\bf G}({\bf k},j){\bf t_k}.
\end{equation}
This quantity and its self-convolution
\begin{equation*}
{\cal T}_{i_1i'_1i_2i'_2}({\bf k},j,j')=\frac{1}{N}\sum_{\bf k'}\theta_{i_1i'_1}({\bf k+k'},j) \theta_{i_2i'_2}({\bf k'},j')
\end{equation*}
appear in (\ref{K}). $C^{(2)a}$ and $C^{(2)s}$ are antisymmetrized and sym\-met\-ri\-zed over spin indices combinations of se\-cond-order cumulants. Due to the symmetry of the problem, $C^{(1)}$, $C^{(2)a}$, and $C^{(2)s}$ do not depend on spin indices of electron operators forming these cumulants. In Eq.~(\ref{K}), quantities $V^a$ and $V^s$ are reducible four-leg vertices, which are analogously antisymmetrized and symmetrized over spin indices. $V^a$  satisfies the fol\-lo\-wing Bethe-Salpeter equation:
\begin{eqnarray}\label{BSE}
&&V^{a}_{i'ii'_1i_1}({\bf k},j+\nu,j,j',j'+\nu)\nonumber\\
&&\quad=C^{(2)a}_{i'ii'_1i_1}(j+\nu,j,j',j'+\nu)\nonumber\\
&&\quad +T\sum_{\nu'i_2i'_2}\sum_{i_3i'_3}C^{(2)a}_{i'i_3i'_2i_1}(j+\nu,j+\nu',j'+\nu',j'+\nu) \nonumber\\
&&\quad\times{\cal T}_{i_3i'_3i_2i'_2}({\bf k},j+\nu',j'+\nu')\nonumber\\
&&\quad\times V^{a}_{i'_3ii'_1i_2}({\bf k},j+\nu',j,j',j'+\nu').
\end{eqnarray}
The equation for $V^s$ looks similarly except that $C^{(2)a}$ is substituted with $C^{(2)s}$. $N$ is the number of sites.

The thermodynamic averaging and operator time dependencies in cumulants are determined by the site Hamiltonian (\ref{local}). Therefore, it is convenient to perform the calculation of cumulants in the re\-pre\-sen\-ta\-ti\-on of its eigenvectors $|\lambda\rangle$. For the first-order cumulant, this gives
\begin{eqnarray}\label{C1}
C^{(1)}_{i'i}(j)&=&\frac{1}{Z}\sum_{\lambda\lambda'}
\frac{{\rm e}^{-\beta E_\lambda}+{\rm e}^{-\beta E_{\lambda'}}}{{\rm i}\omega_j+E_\lambda-E_{\lambda'}}\nonumber\\ &&\times\langle\lambda|a_{i\sigma}|\lambda'\rangle\langle\lambda'|a^\dagger_{i'\sigma}
|\lambda\rangle,
\end{eqnarray}
where $E_\lambda$ is the eigenenergy corresponding to $|\lambda\rangle$ and the partition function $Z=\sum_\lambda\exp(-\beta E_\lambda)$ with $\beta=1/T$. The expression for the second-order cumulant is much more complicated and can be found in \cite{Sherman20a}.

Equations (\ref{Larkin})--(\ref{C1}) and the expression for the se\-cond-order cumulant form a closed set, which can be solved by iteration for given values of $U$, $J$, $T$, and $\mu$ expressed in units of $t$. The calculation consists of two stages. In the first stage, for an electron Green's function obtained in the previous step, the Bethe-Salpeter equations (\ref{BSE}) for $V^a$ and the analogous equation for $V^s$ are solved. As starting values, the respective second-order cumulants $C^{(2)a}$ and $C^{(2)s}$ are used. This calculation stage is significantly simplified by observing that the matrix index of the linear system (\ref{BSE}) consists of only three variables -- $i'$, $i_1$, and $\nu$, while other variables -- ${\bf k}$, $j$, $j'$, $i$, and $i'_1$ -- are parameters. For the considered parameters, 7--10 iteration steps were enough to achieve convergence. In the second stage, the obtained vertices are used for calculating Green's function from equations~(\ref{K}) and (\ref{Larkin}). It is used for obtaining new vertices, and this cycle is repeated until convergence. For the considered parameters, 20--40 cycles were necessary for this. As the initial Green's function, we use that, which is obtained from ${\bf K}$ approximated by the first term in (\ref{K}) -- the first-order cumulant. No artificial broadening was introduced. The integration over wave vectors was approximated by the summation over the mesh of an 8$\times$8 lattice. It has nothing to do with the crystal finiteness, rather it is an approximate method of numerical integration, and the obtained results correspond to an infinite system. For low temperatures, $T\lesssim 0.1t$, the system is close to the transition to the long-range antiferromagnetic ordering. In these conditions, the convergence of the iteration procedure deteriorates. This problem is connected with terms in $C^{(2)}$ \cite{Sherman20a} having the mul\-ti\-pli\-er $1/T$. To stabilize the iteration procedure, we substituted this multiplier by $1/(T+\xi)$ with $\xi=0.2t$ in all such terms. Small variations of $\xi$ produce minor changes in spectral function shapes and electron concentrations.

\section{Densities of states}
\begin{figure*}[htb]
\centerline{\resizebox{1.5\columnwidth}{!}{\includegraphics{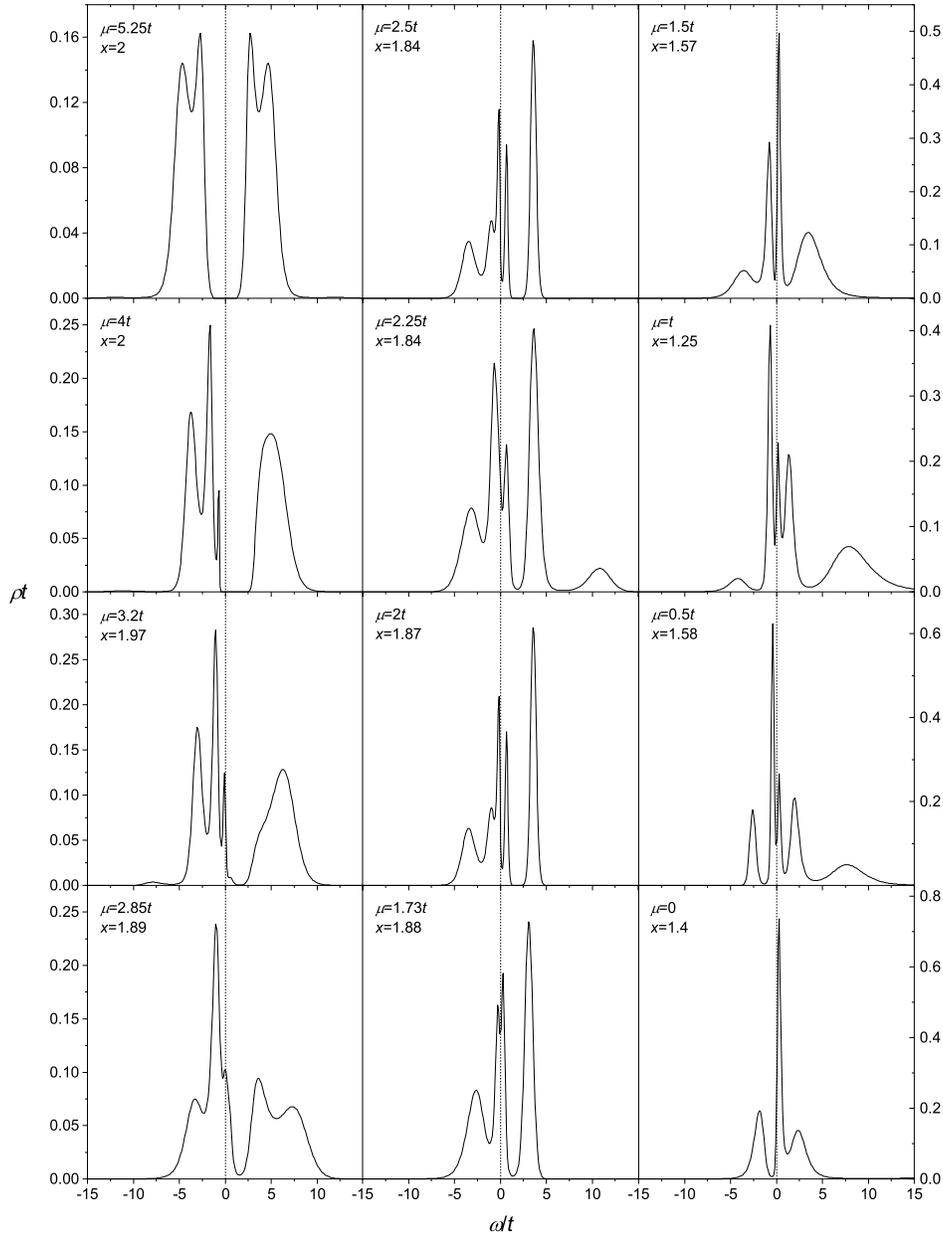}}}
\caption{The evolution of the DOS with the change of the chemical potential. $U=6t$, $J=1.5t$, and $T=0.13t$. Values of $\mu$ and respective electron concentrations are indicated in panels.} \label{Fig2}
\end{figure*}
The DOSs, $\rho_1(\omega)=-(\pi N)^{-1}\sum_{\bf k}{\rm Im}\,G_{11}({\bf k},\omega)$, ob\-tai\-ned in the above iteration procedure for the set of parameters $U=6t$, $J=1.5t$, and $T\approx 0.13t$ are shown in figure~\ref{Fig2}. Due to the symmetry of the problem, $G_{i,-i}=0$ and $G_{11}=G_{-1,-1}$. Therefore, the DOS for only one orbital is shown in the figure. In carrying out the analytic continuation from the imaginary to real frequency axis the maximum entropy method \cite{Press,Jarrell,Habershon} was used. The DOS at half-filling $\mu_{\rm hf}=5.25t$ has the shape typical for the insulating state with the broad Mott gap centered at the Fermi level. As in the one-band Hubbard model, at low temperatures, the spectrum has the four-band structure \cite{Preus,Grober} due to pronounced dips at $\omega\approx\pm 3.75t$. The mechanism of their formation is the same as in the one-band model -- the reabsorption of electrons on transfer frequencies of the site Hamiltonian \cite{Sherman18}. For the considered model with the above parameters, near half-filling, these are $E_{31}-E_{21}=2U-2J-\mu$ and $E_{21}-E_{11}=U-3J-\mu$. Here $E_{n1}$ are the lowest eigenenergies of the site Hamiltonian (\ref{local}) with $n$ electrons.

On the other hand, there is an essential difference between widths of Mott gaps in the present and one-band models at half-filling. In the latter model, the point $U=6t$, $T=0.13t$ is near the boundary of the Mott metal-insulator transition, and, therefore, the gap is small. As seen in figure~\ref{Fig2}, in the HK model, the gap is large and equals approximately $3t$. Below, considering phase diagrams, we shall see that setting $J=0$ shrinks the gap width to a small value, which is close to that in the one-band model. Thus, a finite Hund coupling increases the Mott gap, which can be considered as correlation strengthening. A similar dependence of the gap width on $J$ was earlier found in the half-filled HK model in the DMFT calculations \cite{Georges13}.

Figure~\ref{Fig2} shows also that, with doping, there appears a sharp peak at or in the nearest vicinity of the Fermi level. Similar peaks were found in the spectra of the one- and two-band Hubbard models and were interpreted as spin-polaron peaks -- manifestations of bound states of electrons and spin excitations \cite{Sherman19b,Sherman20a}. This interpretation is supported by the character of diagrams leading to these peaks and by their similarity with spectral manifestations of spin-polaron peaks in the $t$-$J$ model \cite{Schmitt,Ramsak,Sherman94}. An additional argument in favor of this interpretation is the disappearance of the peaks with a temperature elevation when spin excitations are damped.
\begin{figure}[thb]
\centerline{\resizebox{0.99\columnwidth}{!}{\includegraphics{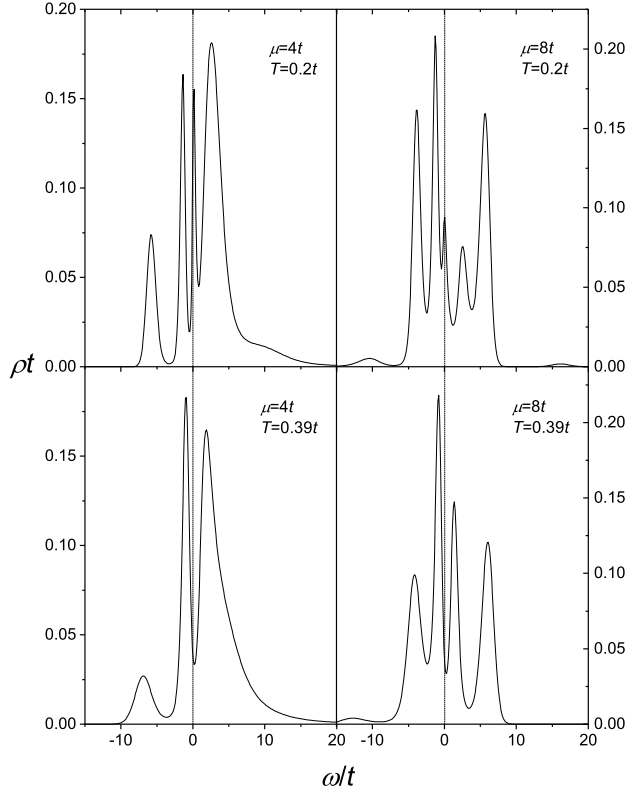}}}
\caption{The evolution of the DOS with the change of the temperature. $U=6t$ and $J=0$. Other parameters are indicated in panels. For both values of $\mu$, temperature lowering leads to the appearance of the spin-polaron peak at the Fermi level.} \label{Fig3}
\end{figure}
Figure~\ref{Fig3} illustrates this statement.

\section{Negative electron compressibility}
Electron concentrations
\begin{equation*}
x=4\int_{-\infty}^{\infty}\frac{\rho_1(\omega){\rm d}\omega}{\exp(\beta\omega)+1},
\end{equation*}
corresponding to DOSs in figure~\ref{Fig2} are shown in the respective panels. Here factor 4 takes into account two values of spin projection and two orbitals. As can be seen, the dependence $x(\mu)$ is not monotonic.
\begin{figure}[thb]
\centerline{\resizebox{0.99\columnwidth}{!}{\includegraphics{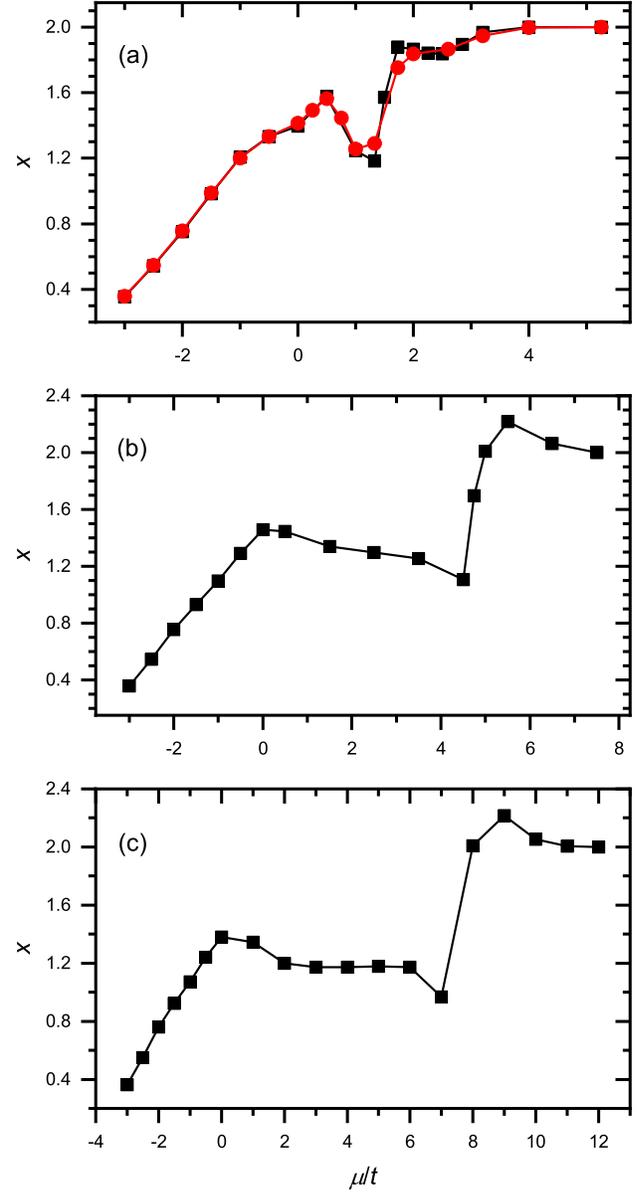}}}
\caption{Dependencies $x(\mu)$ for $U=6t$, $J=1.5t$, $T=0.13t$ (black squares and line) and $T=0.2t$ (red circles and line) (a); $U=5t$, $J=0$, and $T=0.25t$ (b); $U=8t$, $J=0$, and $T=0.26t$ (c).} \label{Fig4}
\end{figure}
It is shown in figure~\ref{Fig4}(a), which demonstrates two NEC regions near $\mu=0$ and $\mu=U-3J=1.5t$. In this figure, the dependencies $x(\mu)$ are shown in the range from $\mu=-3t$ to the chemical potential of half-filling $\mu_{\rm hf}$. As mentioned above, the dependence for larger $\mu$ is obtained by the inversion of this curve in the point $x=2$, $\mu=\mu_{\rm hf}$. Thus, two other NEC regions are located in the range $\mu>\mu_{\rm hf}$. As follows from figure~\ref{Fig4}(a), the temperature increase smooths out the dependence $x(\mu)$ in the NEC regions.

These regions have their origin in the strong dependence of cumulants on $\mu$. The eigenenergies $E_\lambda$ in (\ref{C1}) depend on $\mu$ through the term $-\mu n$, where $n$ is the number of electrons in the state $|\lambda\rangle$. Therefore, as $\mu$ varies, states with different $n$ become alternately the ground state of the Hamiltonian (\ref{HK}). Due to the Boltzmann factors $\exp(-\beta E_\lambda)$ in (\ref{C1}) and in higher-order cumulants, at low temperatures, this state (or states, in the case of degeneracy) and states obtained from it (them) by the creation or annihilation of an electron make the main contribution to the cumulant and, through equations (\ref{Larkin})--(\ref{C1}), to spectral functions. The spectrum is changed drastically when $\mu$ transfers between regions with different ground states, and these changes occur in narrow ranges of $\mu$ at low $T$. The sharp variations of spectra lead to the strong dependence $x(\mu)$ and the appearance of the NEC regions. We notice that the description of Green's function in terms of cumulants strongly depending on $\mu$ is the consequence of the power expansion around the atomic limit, which, in its turn, follows from strong electron correlations. Hence the NEC regions are distinctive features of such correlations.

\begin{figure}[thb]
\centerline{\resizebox{0.99\columnwidth}{!}{\includegraphics{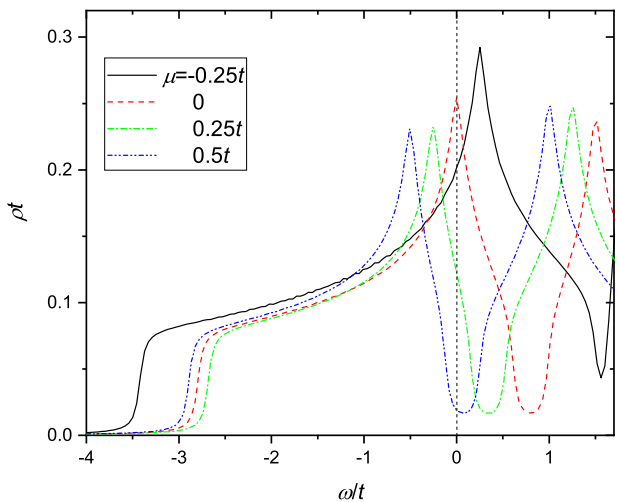}}}
\caption{Densities of states in the first-order approximation. $U=6t$, $J=1.5t$, $T=0.1t$, and $\mu=-0.25t$ (black solid line, $x=1.58$), 0 (red dashed line, $x=1.3$), $0.25t$ (green dash-dotted line, $x=1.24$), and $0.5t$ (blue dash-dot-dotted line, $x=1.29$).} \label{Fig5}
\end{figure}
The mentioned sharp changes of DOSs are seen already in the first order of the SCDT expansion. Figure~\ref{Fig5} demonstrates these changes in more detail for $\mu\approx 0$. Since $C^{(1)}$ in (\ref{C1}) has no imaginary part on the real axis, here we use a 160$\times$160 lattice and the artificial broadening $\eta=0.04t$. For $\mu=-0.25t$, the DOS has a broad band peaked near $\omega=0.25t$ (the black solid curve). With increasing $\mu$, the band does not shift to lower frequencies, as it would happen in the case of rigid bands. Due to the band-structure reconstruction caused by the atomic ground-state change, the band shrinks drastically, and its lower edge shifts to higher frequencies (the red dashed curve for $\mu=0$). As a consequence, the concentration defined by the area under the curve in the range $\omega<0$ decreases sharply. With further increase of the chemical potential, the band continues to shrink, and $x$ decreases (the green dash-dotted curve for $\mu=0.25t$). The electron concentration starts to grow only at $\mu=0.5t$ (the blue dash-dot-dotted curve).

There are five regions of $\mu$ with different values of $n$ in the ground state of the Hamiltonian (\ref{local}). Boundaries between these regions are located near $\mu=0$, $U-3J$, $2U-2J$, and $3U-5J$. Sharp changes of electron bands and NEC regions appear just around these boundaries. The NECs near the two lowest boundaries are seen in figure~\ref{Fig4}(a). In the one-band Hubbard model, there are only two such boundaries, and, at low $T$, NECs are observed near these boundaries \cite{Sherman20b}. In figure~\ref{Fig4}, the dependence $x(\mu)$ is shown for two other sets of parameters also. Two NEC regions near boundary values of $\mu$ are seen for these parameters as well. In the single-site DMFT, for $J=0$, the dependence $x(\mu)$ has plateaus at integer values of $x$ \cite{Rozenberg}. As seen in figure~\ref{Fig4}(b), for moderate repulsions, interactions of electrons with fluctuations transform the plateau at $x=1$ into an extended NEC. In other words, the energy gap responsible for the plateau is filled by excitations, which arise due to the interactions. At larger repulsions, some part of the plateau and the respective energy gap are retained (see figure~\ref{Fig4}(c)). However, it is located at $x$ somewhat larger than 1. The reason for this deviation is the fact that the gap associated with the plateau separates bands rather than atomic levels. For the plateau was located exactly at $x=1$, the two bands should be symmetric, just as in the particle-hole symmetric one-band Hubbard model. In the considered model, such symmetry is absent, which results in the mentioned shift of the plateau from the $x=1$ position.

\begin{figure}[thb]
\centerline{\resizebox{0.99\columnwidth}{!}{\includegraphics{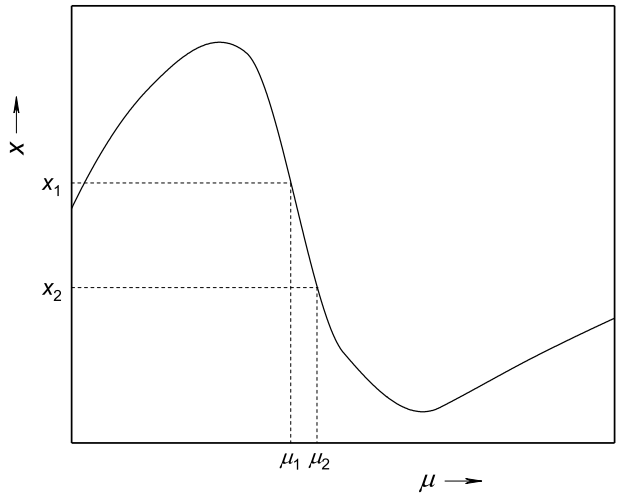}}}
\caption{The dependence $x(\mu)$ in the region of negative electron compressibility.} \label{Fig6}
\end{figure}
If the considered electronic system does not in\-te\-ract with other systems, it remains indefinitely in a state with a chosen value of $\mu$, even if it lies in a NEC region. However, if there is another system, which can exchange energy with the electron system, NEC regions lead to charge instability. Indeed, let us suppose that this system consists of two parts with slightly different electron concentrations and chemical potentials. The values of these $x$ and $\mu$ are located in a NEC region, as shown in figure~\ref{Fig6}. Since $\mu_2>\mu_1$, it is energetically favorable to transfer an electron from the second part ($\mu_2$, $x_2$) to the first part ($\mu_1<\mu_2$, $x_1>x_2$). With the existence of the energy-absorbing system, such a transition occurs, and the difference in concentrations between two parts grows further. This process of charge separation continues until $x$ in the first part reaches the largest concentration in the NEC, and the second part attains its smallest concentration. The character of the dependence $x(\mu)$ prohibits further separation. Phonons are the first candidate for the mentioned energy-absorbing system. The electron phase separation produces the related segregation in the phonon system since lattice distortions and, due to anharmonicity, phonon spectra will differ in the electron-reach and electron-poor parts (we suppose these parts are large enough to speak about electron concentrations and phonon spectra in them).

In models of strongly correlated electrons, phase-separated ground states were found in many works (see, e.g., \cite{Chang,Heiselberg,White}). Optimization procedures used in them do not need in an auxiliary energy-absorbing system, and the phase separation is obtained in a purely electronic problem. These works were inspired by experimental results in cuprate perovskites, transition metal oxides, and some other classes of crystals, for which phase separation is inherent pro\-per\-ty \cite{Shenoy,Dai,Lang,Sigmund}. Charge instabilities connected with NEC and their manifestations in peculiarities of some other systems were discussed in works \cite{Bello,Kravchenko,Eisenstein,Schakel,Skinner,Riley,He,Dezi}.

No negative compressibility was observed in the HK model on a lattice of infinite connectivity in \cite{Stadler}. The application of the one-site DMFT used in the latter work to models on finite-connectivity lattices implies neglecting an infinite sequence of terms in the effective one-site action \cite{Georges96}. In the omitted part, starting from quartic terms, all orders of creation and annihilation operators of the reference site are contained. Notice that the mentioned quartic terms give corrections to the on-site Coulomb repulsion. In our considered diagrams, some processes described by omitted terms are taken into consideration. On the other hand, our approach does not consider all diagrams also. Therefore, distinctions between our results and outcomes of \cite{Stadler} can be expected.

\section{Phase diagram at half-filling}
\begin{figure}[b]
\centerline{\resizebox{0.99\columnwidth}{!}{\includegraphics{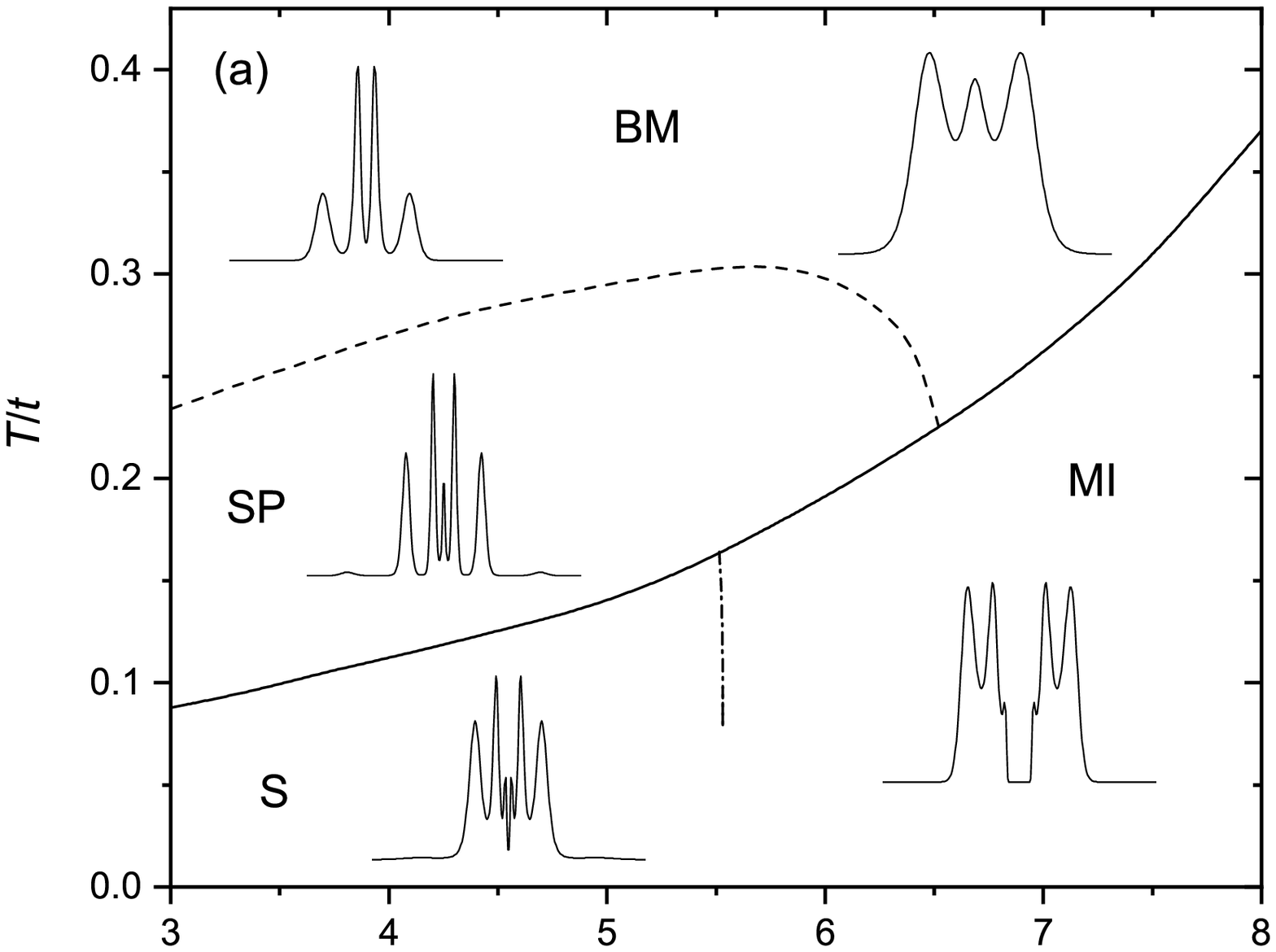}}}
\vspace{2ex}
\hspace{0.3em}\centerline{\resizebox{0.96\columnwidth}{!}{\includegraphics{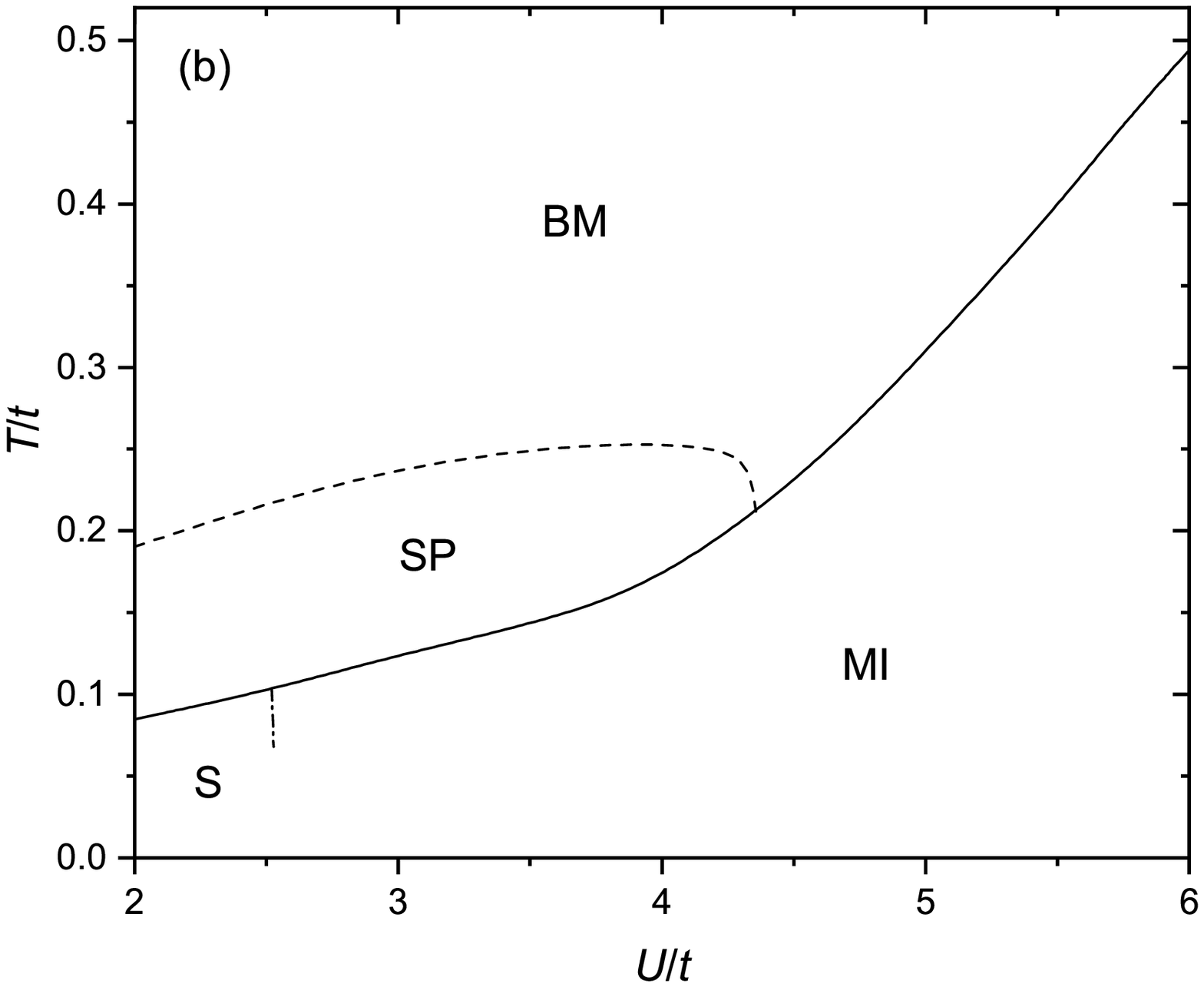}}}
\caption{The phase diagram at half-filling for $J=0$ (a) and $J=U/4$ (b). BM is the abbreviation for the bad-metal state, SP means the state with the spin-polaron peak at the Fermi level, MI is for the Mott insulator, and S denotes the Slater insulating state. Characteristic shapes of DOSs are shown in the respective regions in panel (a).} \label{Fig7}
\end{figure}
Figure~\ref{Fig7} demonstrates different shapes of DOSs ob\-ser\-ved at half-filling in our calculations and the respective regions of parameters. Two panels of the figure allow one to see the influence of the Hund coupling on spectra. In both considered cases, $J=0$ and $J=U/4$, the low-temperature and large-repulsion part of the diagram is occupied by states inherent in the Mott insulator (the abbreviation MI in the diagrams). In these states, around the Fermi level, the DOS features a gap with the width decreasing with $U$. For smaller $U$ and at low $T$, the width of the gap becomes very small and ceases to decrease as $U$ is reduced. Since the considered system shows the short-range antiferromagnetic order (see the next section), we relate this state to the Slater insulator \cite{Slater} (the abbreviation S on the diagram). Comparing panels (a) and (b) of figure~\ref{Fig7}, we see that the Hund coupling extends considerably the MI region at the expense of the S domain.

With increasing temperature, there appears a finite DOS in the gaps. For moderate repulsions, near solid lines in figure~\ref{Fig7}, the produced dip shallows, and, with a further increase of $T$, a narrow peak arises on the Fermi level. It is the spin-polaron peak similar to those discussed above \cite{Sherman19a,Sherman19b}. As mentioned, the peak is a manifestation of the bound states of electrons and spin excitations. Hence, the antiferromagnetic ordering leads to two diametrally opposite occurrences for different temperature ranges at moderate repulsions. At lower temperatures, it is the Slater gap arising due to the scattering of electrons on the antiferromagnetic background. For somewhat higher $T$, there appears the Fermi-level peak revealing bound states of electrons and spin excitations. The region of DOSs with the spin-polaron peak is denoted as SP on phase diagrams of figure~\ref{Fig7}.

At even higher temperatures, DOSs acquire sha\-pes with a finite $\rho(\omega=0)$ without a maximum at the Fermi level or with a broad maximum (for $T\gtrsim t$), as shown in the region denoted as BM in figure~\ref{Fig7}(a). These are bad-metal states.

In this work, we did not consider the character of transitions between states of different types. With the procedure used for calculating real-frequency DOSs -- the maximum entropy method -- it is difficult. As seen from the comparison of two panels of figure~\ref{Fig7}, for a fixed $T$, the Hund coupling decreases significantly the value of the critical repulsion $U_c$ separating metallic and insulating states (solid lines in the figure). This conclusion is in qualitative agreement with the result obtained by DMFT \cite{Georges13}. For $J=0$, the dependence $U_c(T)$ is similar to that calculated in the one-band Hubbard model for strong repulsions in \cite{Sherman18} and mo\-de\-ra\-te $U$ in \cite{Schafer}.

\section{Susceptibilities}
To investigate possible orderings in the model, we consider spin $\chi^{\rm sp}$, charge $\chi^{\rm ch}$, and two orbital sus\-cep\-ti\-bi\-li\-ti\-es $\chi^{\rm orb}_{+-}$ and $\chi^{\rm orb}_{zz}$. For the considered two-band Hamiltonian, there is no rotation invariance for the orbital moment at $J\neq 0$ \cite{Georges13}. Therefore, the two orbital susceptibilities may differ. All these susceptibilities can be expressed through the above-calculated Green's function and vertices. For example, the spin sus\-cep\-ti\-bi\-li\-ty reads
\begin{eqnarray}\label{chi_sp}
&&\chi^{\rm sp}({\bf k},\nu)=\frac{1}{2}\sum_{i'i}\big\langle\big\langle S^{+1}_{{\bf k}i'}|S^{-1}_{{\bf k}i}\big\rangle\big\rangle\nonumber\\
&&\quad=-\frac{T}{N}\sum_{{\bf k'}j}G_{11}({\bf k+k'},\nu+j)G_{11}({\bf k'},j) \nonumber\\
&&\quad-T^2\sum_{jj'}F({\bf k},\nu+j,j)F({\bf k},\nu+j',j')\nonumber\\
&&\quad\times\Big[V^a_{1111}({\bf k},j+\nu,j'+\nu,j',j)\nonumber\\
&&\quad+V^a_{1,-1,-1,1}({\bf k},j+\nu,j'+\nu,j',j)\Big].
\end{eqnarray}
Susceptibilities
\begin{eqnarray*}
&&\chi^{\rm ch}({\bf k},\nu)=\frac{1}{4}\sum_{\sigma'\sigma}\sum_{i'i} \big\langle\big\langle\big(n_{{\bf k}i'\sigma'}-\langle n_{{\bf k}i'\sigma'}\rangle\big)\big|\\
&&\quad\quad\times\big(n_{{\bf k}i\sigma}-\langle n_{{\bf k}i\sigma}\rangle\big)\big\rangle\big\rangle,\\
&&\chi^{\rm orb}_{+-}({\bf k},\nu)=\frac{1}{2}\sum_{\sigma'\sigma}\big\langle\big\langle L^{+1}_{{\bf k}\sigma'}\big|L^{-1}_{{\bf k}\sigma}\big\rangle\big\rangle,{\rm  and}\\
&&\chi^{\rm orb}_{zz}({\bf k},\nu)=\sum_{\sigma'\sigma}\big\langle\big\langle L^z_{{\bf k}\sigma'}\big|L^z_{{\bf k}\sigma}\big\rangle\big\rangle
\end{eqnarray*}
are expressed by similar equations, in which the sum of vertices in the square brackets in (\ref{chi_sp}) is substituted by
\begin{eqnarray*}
&& V^s_{1111}({\bf k},j+\nu,j'+\nu,j',j)\\
&&\quad+V^s_{1,-1,-1,1}({\bf k},j+\nu,j'+\nu,j',j),\\
&&V^s_{11,-1,-1}({\bf k},j+\nu,j'+\nu,j',j),{\rm  and}\\
&&V^s_{1111}({\bf k},j+\nu,j'+\nu,j',j)\nonumber\\
&&\quad-V^s_{1,-1,-1,1}({\bf k},j+\nu,j'+\nu,j',j),
\end{eqnarray*}
respectively. In the above equations, $\big\langle\big\langle{\cal O}\big|{\cal O}^\dagger \big\rangle\big\rangle$ is Green's function of the respective operators, $S^\sigma_{{\bf k}i}$ is the Fourier transform of the spin operator $S^\sigma_{{\bf l}i}=a^\dagger_{{\bf l}i\sigma}a_{{\bf l}i,-\sigma}$, $L^i_{{\bf k}\sigma}$ and $L^z_{{\bf k}\sigma}$ are Fourier transforms of the orbital operators $L^i_{{\bf l}\sigma}=a^\dagger_{{\bf l}i\sigma}a_{{\bf l},-i,\sigma}$ and $L^z_{{\bf l}\sigma}=\frac{1}{2}\sum_i i a^\dagger_{{\bf l}i\sigma}a_{{\bf l}i\sigma}$,
\begin{equation*}
F({\bf k},j,j')=\frac{1}{N}\sum_{\bf k'}\Pi({\bf k+k'},j)\Pi({\bf k'},j'),
\end{equation*}
and $\Pi({\bf k},j)=1+t_{11}({\bf k})G_{11}({{\bf k},j})$.

\begin{figure*}[t]
\centerline{\resizebox{1.7\columnwidth}{!}{\includegraphics{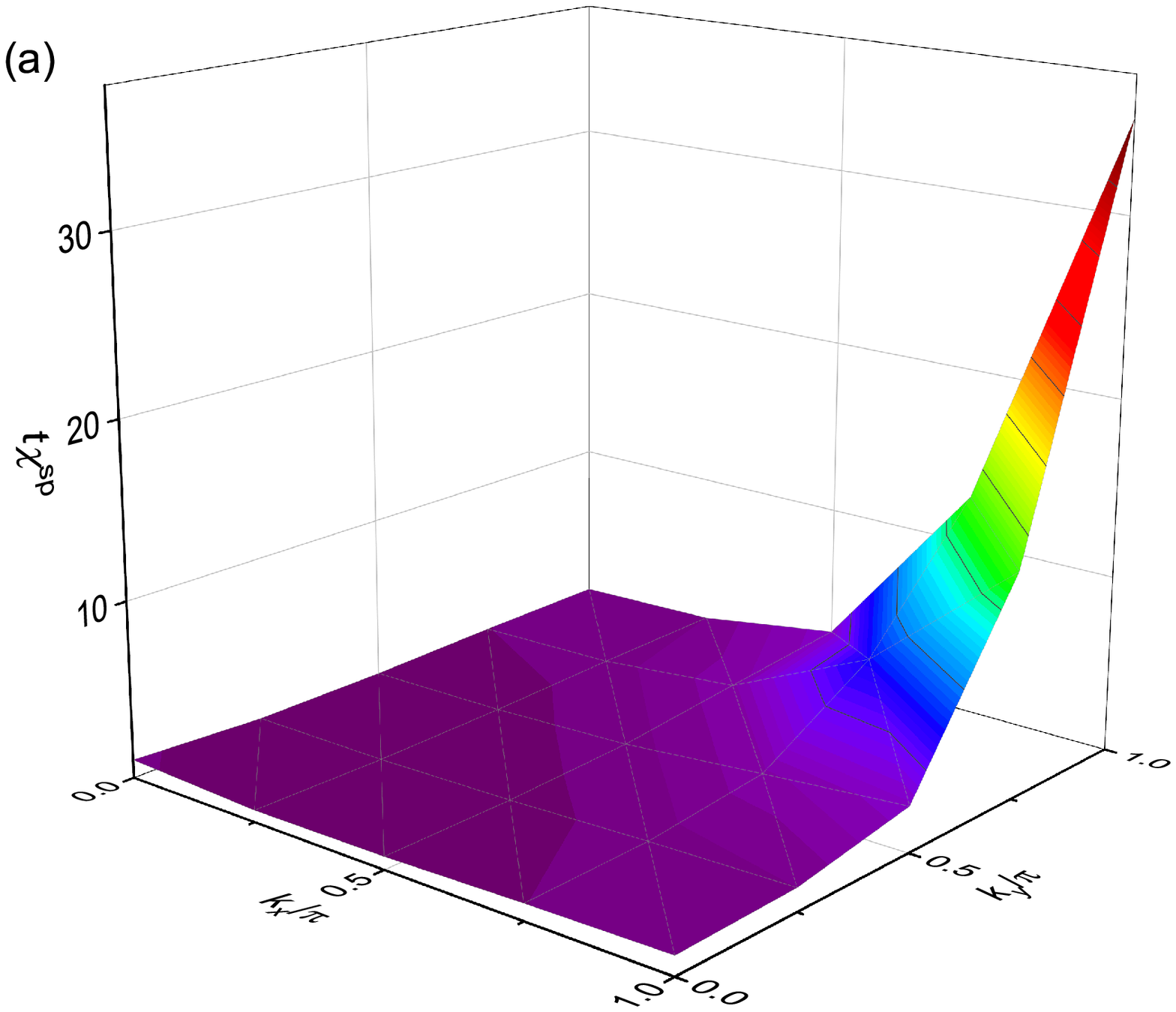}\hspace{5em} \includegraphics{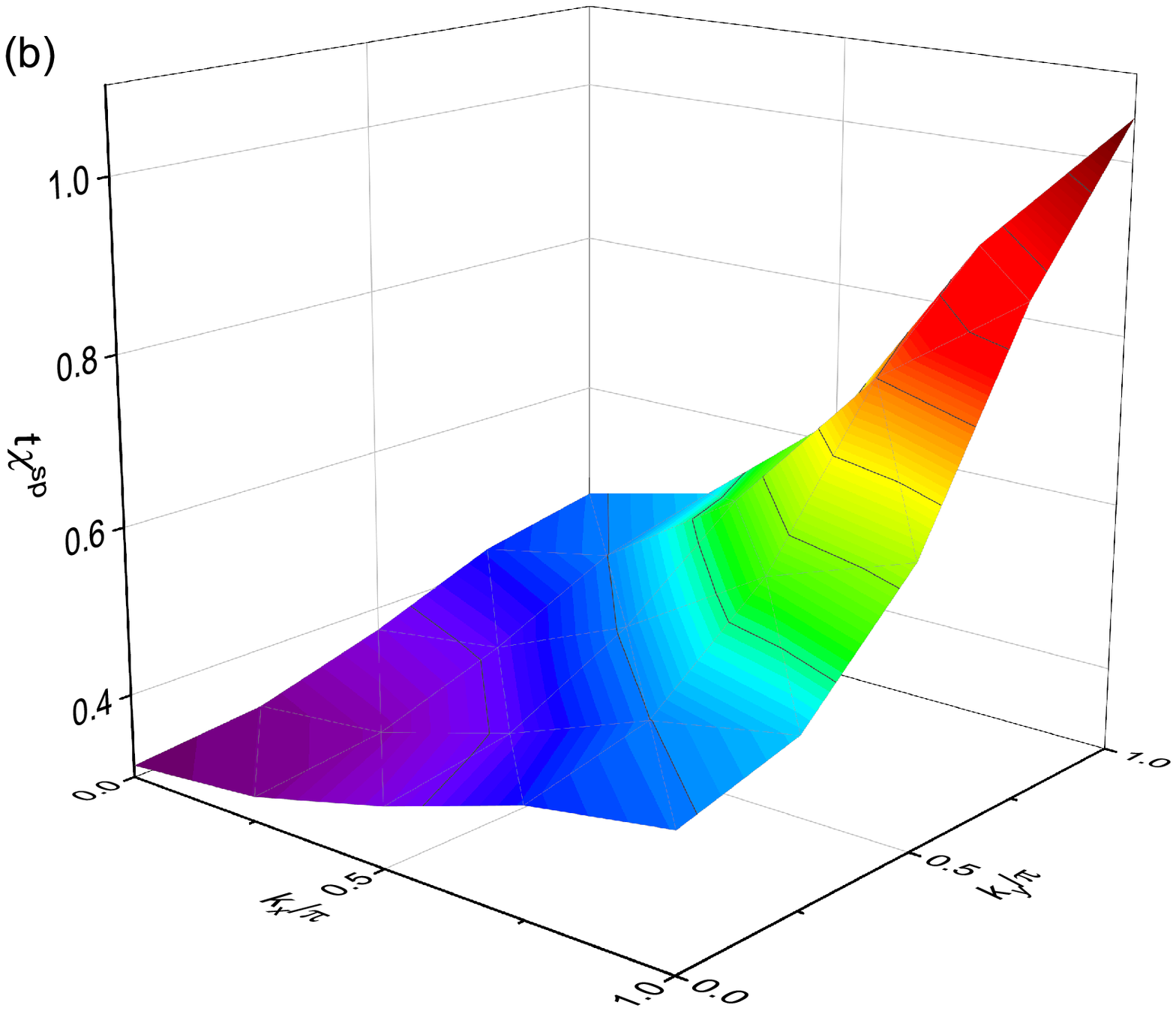}}}
\vspace{3ex}
\centerline{\resizebox{1.7\columnwidth}{!}{\includegraphics{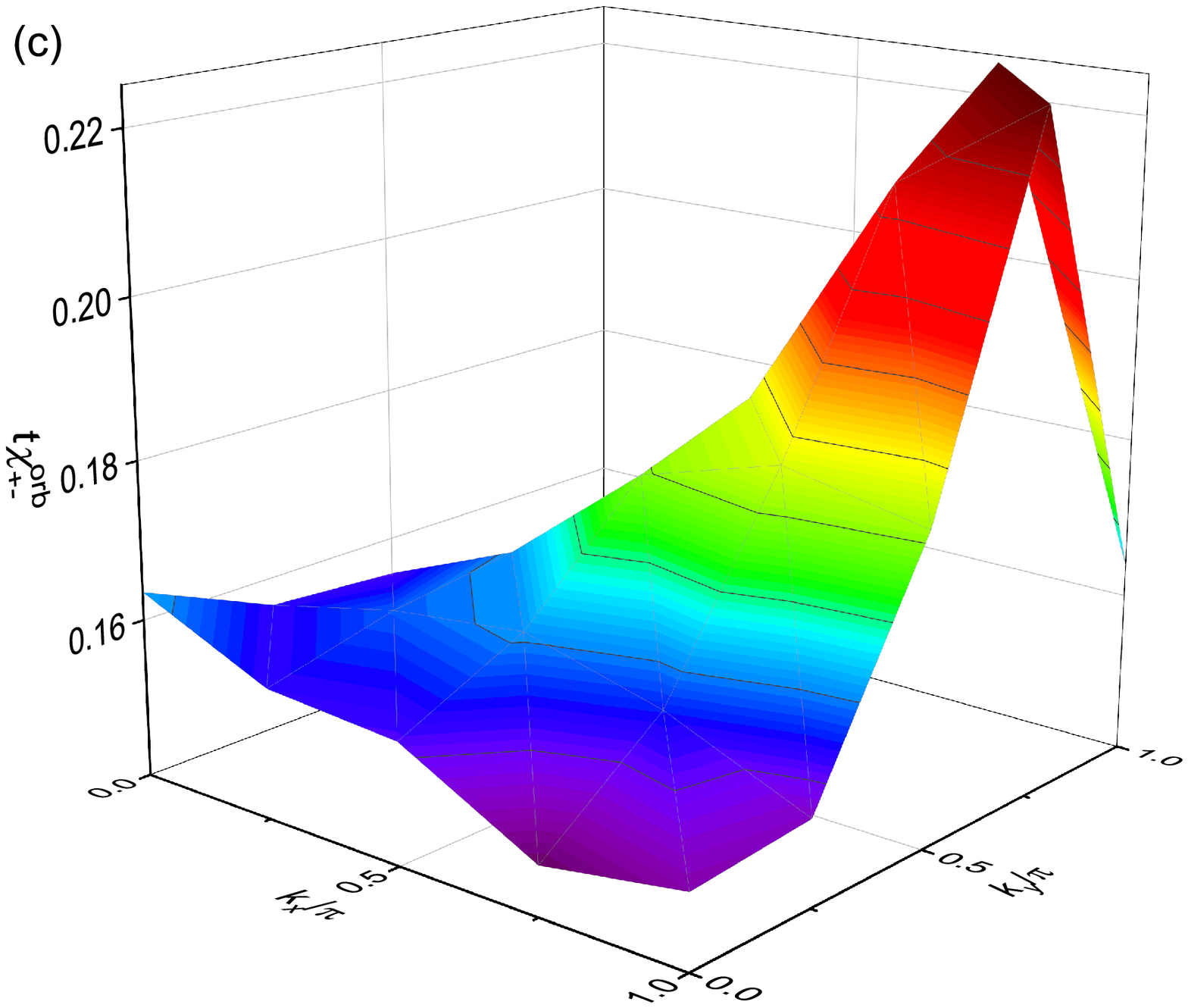}\hspace{5em} \includegraphics{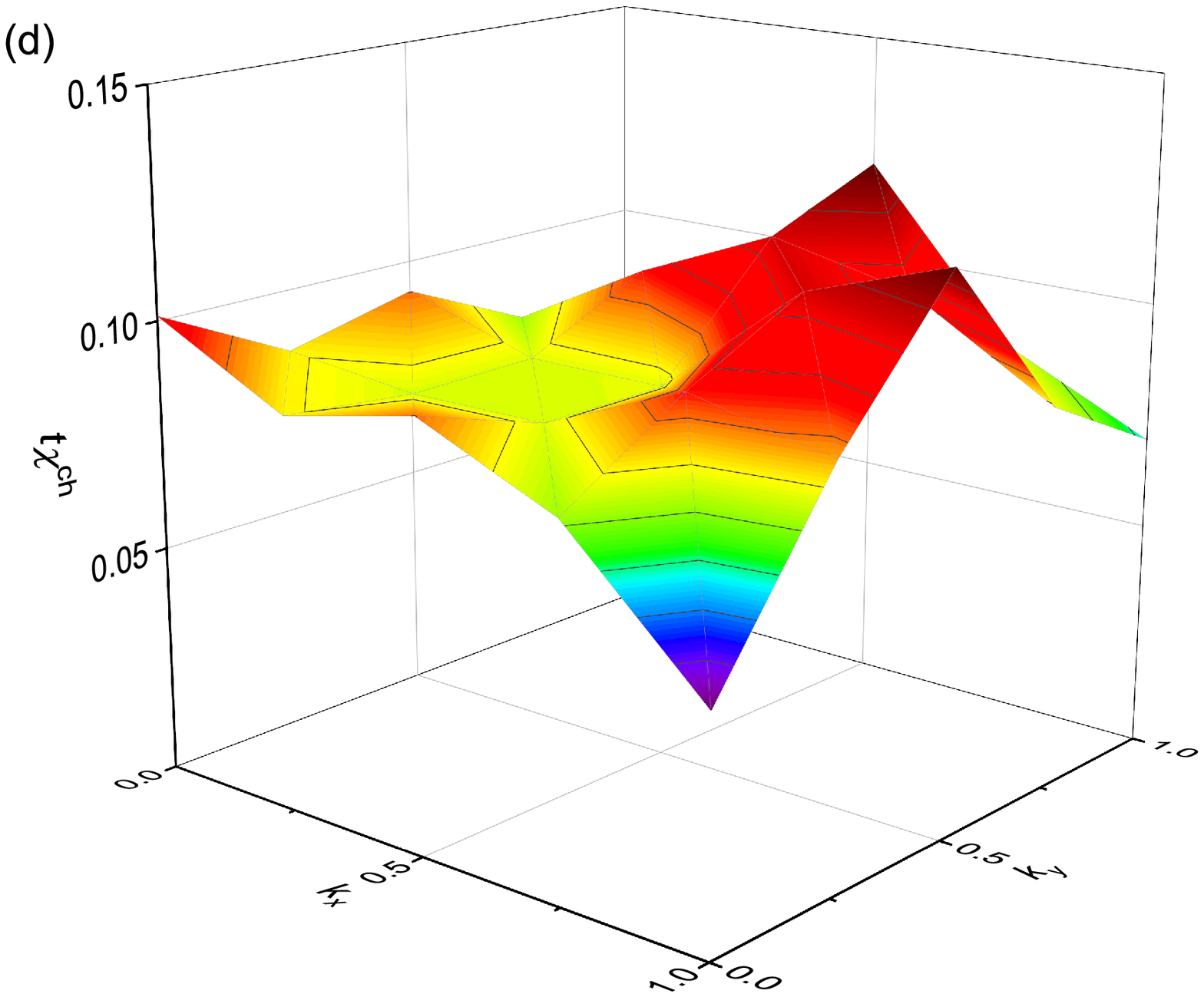}}}
\caption{Zero-frequency susceptibilities in the first quadrant of the Brillouin zone for $U=6t$ and $T=0.13t$. Spin susceptibilities for $J=1.5t$, $\mu=\mu_{\rm hf}=5.25t$ (a) and $J=0$, $\mu=\mu_{\rm hf}=9t$ (b). The orbital susceptibility $\chi^{\rm orb}_{+-}$ for $J=1.5t$, $\mu=1.5t$ (c) and the charge susceptibility for $J=1.5t$, $\mu=-t$ (d).} \label{Fig8}
\end{figure*}
To investigate possible orderings, we consider zero-frequency susceptibilities. They change significantly as the chemical potential transfers between regions with different atomic ground states. For $U=6t$ and $J=1.5t$, in the $\mu$ range from $\mu_{\rm hf}=5.25t$ to the vicinity of the first NEC at $\mu\approx 2.5t$, the spin susceptibility is large at the antiferromagnetic wave vector ${\bf Q}=(\pi,\pi)$ (see figure~\ref{Fig8}(a)). Such a value of $\chi^{\rm sp}$ is connected with the formation of the coupled spin $S=1$ of two electrons owing to Hund's coupling. For $J=0$, when the coupled spin does not arise, $\chi^{\rm sp}({\bf Q},0)$ is much smaller (see figure~\ref{Fig8}(b); panels (a) and (b) correspond to half-filling and the same $U$ and $T$). It is comparable to the respective susceptibility in the one-band Hubbard model at similar parameters \cite{Sherman19a}. The formations of coupled spin and orbital moments compete with each other -- for the former moment, two electrons have to occupy different orbitals, while for the latter, they should be on the same orbital. Therefore, in the mentioned range of $\mu$, orbital susceptibilities, as well as the charge susceptibility, are three to four orders of magnitude smaller than $\chi^{\rm sp}({\bf Q},0)$ for $J=1.5t$. For this Hund coupling, in all considered range of parameters, the orbital susceptibilities $\chi^{\rm orb}_{+-}$ and $\chi^{\rm orb}_{zz}$, though not equal, have close values.

For $J=0$, terms of the Hamiltonian (\ref{HK}), which violate the rotation symmetry of the orbital moment components, disappear. As a consequence, $\chi^{\rm orb}_{+-}$ be\-co\-mes equal to $\chi^{\rm orb}_{zz}$. Besides, due to the symmetry of the Hamiltonian for the exchange $\sigma\leftrightarrow i$, the spin susceptibility is equal to the orbital susceptibilities for all values of parameters.

For $U=6t$ and $J=1.5t$, as $\mu$ attains the first NEC at $\mu\approx 2.5t$, the spin susceptibility decreases drastically and becomes comparable to three other susceptibilities. This points to the disruption of the coupled spin. Nevertheless, as for larger values of $\mu$, all susceptibilities peak at {\bf Q}, pointing to the short-range antiferromagnetic ordering. For $J=0$, susceptibilities are at their maxima at this wave vector also. The situation is changed near $\mu=U-3J$, the chemical potential corresponding to the change of the atomic ground state. Both for $J=1.5t$ and $J=0$, in a narrow range around this $\mu$, one or several susceptibilities become incommensurate (see figure~\ref{Fig8}(c)). For smaller chemical potential outside of this range and up to $\mu\approx 0$, all susceptibilities peak again at {\bf Q} and have comparable values. For $U=8t$, $J=0$, and $2t\lesssim\mu\lesssim 6t$, in the range of plateau in figure~\ref{Fig4}(c), they remain practically unchanged.

The last significant modification in the susceptibilities occurs near $\mu=0$, the other chemical potential corresponding to the change of the atomic ground state. For $\mu\lesssim 0$, all susceptibilities become small and show no ordering (see figure~\ref{Fig8}(d)).

\section{Conclusion}
In this work, the two-orbital Hubbard-Kanamori model on a two-dimensional square lattice was investigated. For this purpose, we used the strong-coupling diagram technique. Diagrams taken into account in this work allowed us to consider the influence of interactions of electrons with spin, charge, and orbital fluctuations of all ranges. Intervals of moderate and strong on-site Coulomb repulsions $2t\leq U\leq 8t$ and temperatures $0.07\lesssim T\lesssim t$ for two values of Hund's couplings, $J=U/4$ and $J=0$, were considered. Calculated densities of states show that some properties inherent in the one-band Hubbard model are retained in the present model. In particular, near half-filling, at low temperatures, densities of states have the four-band shapes due to dips at atomic transfer frequencies $2U-2J-\mu$ and $U-3J-\mu$. The dips are caused by the reabsorption of electrons at these frequencies. In the regions of $\mu$ with the pronounced short-range antiferromagnetic ordering and low temperatures, a sharp peak of the density of states is observed at the Fermi level. For strong repulsions, the peak is seen with doping, while for moderate $U$, it appears at half-filling also. The peak is a manifestation of spin polarons -- bound states of electrons and spin excitations. In comparison with the one-band Hubbard model, a finite Hund coupling increases the width of the Mott gap at half-filling. An analogous result was earlier obtained with DMFT.

For $J=0$ and large Coulomb repulsions, the dependence of the electron concentration $x$ on $\mu$ has plateaus near $x=1$ and $x=3$. The plateaus correspond to gaps in the densities of states. With decreasing $U$ or increasing $J$, the gaps are filled by excitations, and the plateaus disappear. In contrast to the DMFT result, in our calculations, the plateaus are somewhat shifted from the positions $x=1$ and $x=3$ due to the asymmetry of bands separated by respective gaps.

At low temperatures, the dependence $x(\mu)$ has four regions of negative electron compressibility. They arise due to sharp changes in cumulants of electron operators occurring near chemical potentials $\mu=0$, $U-3J$, $2U-2J$, and $3U-5J$. At these values of $\mu$, one ground state of the site Hamiltonian gives way to another one characterized by another number of electrons in it. Such a strong influence of site processes on the entire electronic spectrum is a consequence of strong electron correlations, which justify the app\-li\-ca\-ti\-on of the perturbation series expansion around the atomic limit. The regions of negative electron compressibility lead to the phase separation if phonons or some other system absorbs energy released by electrons in the course of segregation into electron-poor and electron-rich regions.

The $T$-$U$ phase diagram at half-filling contains regions of the Mott and Slater insulators, the state with spin-polaron peak, and the bad metal. The Mott-insulator state is characterized by the gap around the Fermi level, which width decreases with $U$. Such a state is observed at large $U$ and small $T$. For smaller repulsions, in the Slater insulating state, the gap is narrower and does not change with $U$. The sharp spin-polaron peak at the Fermi level is inherent in the spin-polaron state observed at somewhat larger temperatures. In\-te\-res\-ting\-ly that the an\-ti\-fer\-ro\-mag\-ne\-tic ordering leads to two diametrally opposite occurrences -- the peak at higher temperatures and the gap at lower $T$. Fi\-nal\-ly, the bad-metal state is located at even higher temperatures and characterized by a dip with a finite density of states at the Fermi level or a broad maximum there. In its shape, the line separating insulating and conducting states is similar to that in the one-band Hubbard model. For a fixed $T$, the value of the critical repulsion on this line decreases as $J$ increases. An analogous result was obtained with DMFT.

Zero-frequency spin $\chi^{\rm sp}$, charge $\chi^{\rm ch}$, and orbital $\chi^{\rm orb}$ susceptibilities are drastically changed when the chemical potential transfers between regions with dif\-fe\-rent atomic ground states. For $J=1.5t$, low $T$, and $\mu$ in the range from half-filling $\mu_{\rm hf}=(3U-5J)/2$ to approximately $U-3J$, the spin susceptibility at the antiferromagnetic momentum ${\bf Q}=(\pi,\pi)$ is large due to the formation of the coupled spin $S=1$. In this region, $\chi^{\rm ch}$ and  $\chi^{\rm orb}$ are suppressed. As the chemical potential approaches the value $U-3J$, $\chi^{\rm sp}$ decreases drastically and becomes comparable to other susceptibilities. In the vicinity of $\mu=U-3J$, one or several susceptibilities go incommensurate. In the range $0<\mu<U-3J$, they peak at {\bf Q} again. For $\mu\lesssim 0$, all susceptibilities get small and show no order.


\section*{References}
\begin{thebibliography}{99}
\bibitem{Kanamori}Kanamori J 1963 {\it Prog. Theor. Phys.} {\bf 30} 275
\bibitem{Georges13}Georges A, de' Medici L and Mravlje J 2013 {\it Annu. Rev. Condens. Matter Phys.} {\bf 4} 137
\bibitem{Shenoy}Shenoy V B and Rao C N R (2008) {\it Phil. Trans. R. Soc.} A {\bf 366} 63
\bibitem{Dai}Dai P, Hu J and Dagotto E 2012 {\it Nat. Phys.} {\bf 8} 709
\bibitem{Lang}Lang G, Grafe H-J, Paar D, Hammerath F, Manthey K, Behr G, Werner J and Buchner B 2010 {\it Phys. Rev. Lett.} {\bf 104} 097001
\bibitem{Sigmund}Sigmund E and Müller K A eds 1994 {\it Phase Separation in Cuprate Superconductors} (Berlin: Springer-Verlag)
\bibitem{Georges96}Georges A, Kotliar G, Krauth W and Rozenberg M 1996 {\it Rev.\ Mod.\ Phys.} {\bf 68} 13
\bibitem{Stadler}Stadler K M, Kotliar G, Weichselbaum A and von Delft J 2019 {\it Ann.\ Phys.\ (N.\ Y.)} {\bf 405} 365
\bibitem{Medici}de' Medici L (2017) {\it Phys.\ Rev.\ Lett.} {\bf 118} 167003
\bibitem{Vladimir}Vladimir M I and Moskalenko V A  1990 {\it Theor.\ Math.\ Phys.} {\bf 82} 301
\bibitem{Metzner}Metzner W 1991 {\it Phys.\ Rev.} B {\bf 43} 8549
\bibitem{Pairault}Pairault S, S\'en\'echal D and Tremblay A-M S 2000 {\it Eur.\ Phys.\ J.} B {\bf 16} 85
\bibitem{Sherman18}Sherman A 2018 {\it J. Phys.: Condens. Matter} {\bf 30} 195601
\bibitem{Sherman19a}Sherman A 2019 {\it Eur. Phys. J.} B {\bf 92} 55
\bibitem{Sherman16}Sherman A 2016 {\it Eur.\ Phys.\ J.} B {\bf 89} 91
\bibitem{Sherman19b}Sherman A 2019 {\it Phys.\ Scr.} {\bf 94} 055802
\bibitem{Sherman20a}Sherman A 2020 Spin and charge fluctuations in the two-band Hubbard model (arXiv:2001.09270)
\bibitem{Damascelli}Damascelli A, Hussain Z and Shen Z-X 2003 {\it Rev.\ Mod.\ Phys.} {\bf 75} 473
\bibitem{Armitage}Armitage N P, Fournier P and Greene R L 2010 {\it Rev.\ Mod.\ Phys.} {\bf 82} 2421
\bibitem{Aron}Aron C and Kotliar G 2015 {\it Phys. Rev.} B {\bf 91} 041110
\bibitem{Kubo}Kubo R 1962 {\it J. Phys. Soc. Jpn.} {\bf 17} 1100
\bibitem{Abrikosov} Abrikosov A A, Gor’kov L P and Dzyaloshinskii I E 1965 {\it Methods of Quantum Field Theory in Statistical Physics} (New York: Pergamon Press)
\bibitem{Press}Press W H, Teukolsky S A, Vetterling W T and Flannery B P 1995 {\it Numerical Recipes in Fortran} (Cambridge: Cambridge University Press) chapter 18
\bibitem{Jarrell}Jarrell M and Gubernatis J E 1996 {\it Phys.\ Rept.} {\bf 269} 133
\bibitem{Habershon}Habershon S, Braams B J and Manolopoulos D E 2007 {\it J.\ Chem.\ Phys.} {\bf 127} 174108
\bibitem{Preus}Preuss R, Hanke W and von der Linden W 1995 {\it Phys. Rev. Lett.} {\bf 75} 1344
\bibitem{Grober}Gröber C, Eder R and Hanke W 2000 {\it Phys. Rev.} B {\bf 62} 4336
\bibitem{Schmitt}Schmitt-Rink S, Varma C M and Ruckenstein A E 1988 {\it Phys. Rev. Lett.} {\bf 60} 2793
\bibitem{Ramsak}Ramšak A and Horsch P 1993 {\it Phys. Rev.} B {\bf 48} 10559
\bibitem{Sherman94}Sherman A and Schreiber M 1994 {\it Phys. Rev.} B {\bf 50} 12887
\bibitem{Sherman20b}Sherman A 2020 {\it Phys.\ Scr.} {\bf 95} 015806
\bibitem{Rozenberg}Rozenberg M J 1997 {\it Phys.\ Rev.} B {\bf 55} R4855
\bibitem{Chang}Chia-Chen Chang and Shiwei Zhang 2008 {\it Phys.\ Rev.} B {\bf 78} 165101
\bibitem{Heiselberg}Heiselberg H 2009 {\it Phys.\ Rev.} A {\bf 79} 063611
\bibitem{White}White S R and Scalapino D J 2015 {\it Phys. Rev.} B {\bf 92} 205112
\bibitem{Bello}Bello M S, Levin E I, B. I. Shklovskrii B I and Efros A L 1981 {\em Sov. Phys. JETP} {\bf 53} 822
\bibitem{Kravchenko}Kravchenko S V, Rinberg D A, Semenchinsky S G and Pudalov V M 1990 {\em Phys. Rev.} B {\bf 42} 3741.
\bibitem{Eisenstein} Eisenstein J P, Pfeiffer L N and West K W 1992 {\em Phys. Rev. Lett.} {\bf 68} 674.
\bibitem{Schakel}Schakel A M J 2001 {\it Phys. Rev.} B {\bf 64} 245101
\bibitem{Skinner}Skinner B, Yu G L, Kretinin A V, Geim A K, Novoselov K S and Shklovskii B I 2013 {\em Phys. Rev.} B {\bf 88} 155417
\bibitem{Riley}Riley J M, Meevasana W, Bawden L, Asakawa M, Ta\-ka\-ya\-ma T, Eknapakul T, Kim T K, Hoesch M, Mo S-K and Takagi H 2015 {\em Nature Nanotechnology} {\bf 10} 1043
\bibitem{He}He J, Hogan T, Mion T R, Hafiz H, He Y, Denlinger J D, Mo S-K, Dhital C, Chen X, Lin Q {\em et al.} 2015 {\em Nature Materials} {\bf 14} 577
\bibitem{Dezi}Dezi G, Scopigno N, Caprara A and Grilli M 2018 {\em Phys. Rev.} B {\bf 98} 214507
\bibitem{Slater}Slater J C 1951 {\it Phys. Rev.} {\bf 82} 538
\bibitem{Schafer}Sch\"afer T, Geles F, Rost D, Rohringer G, Arrigoni E, Held K, Bl\"umer N, Aichhorn M and Toschi A 2015 {\it Phys. Rev.} B {\bf 91} 125109

\end{thebibliography}
\end{document}